\documentclass[journal]{IEEEtran}
\usepackage[table,xcdraw]{xcolor}
\usepackage{adjustbox}
\usepackage[utf8]{inputenc}
\usepackage{graphicx}
\usepackage{amssymb}
\usepackage{enumerate}
\usepackage{mathtools}
\usepackage{longtable}
\usepackage{supertabular}
\usepackage{amsmath}
\usepackage{caption}
\usepackage{subcaption} 
\usepackage{placeins}
\usepackage{booktabs}
\usepackage{color}
\usepackage{physics}
\usepackage{algorithm}
\usepackage{algpseudocode}
\usepackage{cite}
\usepackage[colorinlistoftodos]{todonotes}
\usepackage{gensymb}
\usepackage{flushend}
\usepackage{multicol}
\usepackage{multirow}
\usepackage{mwe}
\graphicspath{ {./Images/} }
\usepackage[T1]{fontenc}
\usepackage{nomencl}
\usepackage{lastpage}
\usepackage{wrapfig}
\usepackage{fancyhdr}
\usepackage{lipsum}
\usepackage{cite}
\usepackage{fancyhdr}
\usepackage{tikz}
\usetikzlibrary{shapes,backgrounds}
\usepackage{verbatim}
\usepackage{url}
\usepackage{enumitem}
\usepackage{dirtytalk}
\usepackage{amsmath,amssymb,amsfonts}
\usepackage{xfrac}
\usepackage{textcomp}
\usepackage{xcolor}
\usepackage{color,soul}
\usepackage{scalerel}
\usepackage{csquotes}
\usepackage{comment}
\usepackage{tikz}
\usetikzlibrary{svg.path}

\definecolor{googleblue}{rgb}{0.259,0.522,0.957}
\definecolor{googlegreen}{rgb}{0.060,0.620,0.350}
\definecolor{googlered}{rgb}{0.859,0.267,0.220}
\definecolor{bleudefrance}{rgb}{0.19,0.55,0.91}

\definecolor{orcidlogocol}{HTML}{A6CE39}
\tikzset{
  orcidlogo/.pic={
    \fill[orcidlogocol] svg{M256,128c0,70.7-57.3,128-128,128C57.3,256,0,198.7,0,128C0,57.3,57.3,0,128,0C198.7,0,256,57.3,256,128z};
    \fill[white] svg{M86.3,186.2H70.9V79.1h15.4v48.4V186.2z}
                 svg{M108.9,79.1h41.6c39.6,0,57,28.3,57,53.6c0,27.5-21.5,53.6-56.8,53.6h-41.8V79.1z M124.3,172.4h24.5c34.9,0,42.9-26.5,42.9-39.7c0-21.5-13.7-39.7-43.7-39.7h-23.7V172.4z}
                 svg{M88.7,56.8c0,5.5-4.5,10.1-10.1,10.1c-5.6,0-10.1-4.6-10.1-10.1c0-5.6,4.5-10.1,10.1-10.1C84.2,46.7,88.7,51.3,88.7,56.8z};
  }
}

\newcommand\orcidicon[1]{\href{https://orcid.org/#1}{\mbox{\scalerel*{
\begin{tikzpicture}[yscale=-1,transform shape]
\pic{orcidlogo};
\end{tikzpicture}
}{|}}}}

\usepackage[bookmarks=false]{hyperref} 
\hypersetup{
    colorlinks=true,
}
\begin{document}
\raggedbottom

\twocolumn
\setcounter{page}{1}

\author{
        \IEEEauthorblockN{Brandon C. Colelough,~$^{\orcidicon{0000-0001-8389-3403}}$}
        
        \IEEEauthorblockA{School of Engineering and Information Technology, University of New South Wales, Australia}
}

\title{Australian Energy Market Operator National Electricity Market Network Optimal Power Flow Modelling}

\maketitle

\begin{abstract}
The HELM algorithm was used in this project to solve the optimal power flow problem introduced by a radial PandaPower network formulated from the data given by AEMO on the NEM network. Large losses were observed in the transmission infrastructure surrounding base-load power plants. These losses were not observed in areas that had a higher percentage of renewable power generation. Furthermore, the voltage levels present at the Hydro and Wind farms across Tasmania and South Australia were found to be stable in their steady state. A distributed network of renewable infrastructure is then proposed as a solution to the issues facing the NEM network. 
\end{abstract} 

\section{Preliminary Information}\label{Preliminary Information}
\makenomenclature
\mbox{}
\nomenclature{$S_k$}{Total power}
\nomenclature{$P_k$}{Real power}
\nomenclature{$Q_k$}{Reactive power}
\nomenclature{$V_k$}{Voltage magnitude}
\nomenclature{$\theta_k$}{Voltage phase}
\nomenclature{$E_s$}{Complex voltage}
\nomenclature{$y_{mk}$}{Transmission admittance elements}
\nomenclature{$Y_k$}{Shunt admittances}
\nomenclature{$f_m$}{set of simultaneous non-linear equations}
\nomenclature{$J$}{Jacobian matrix}
\nomenclature{$Lat$}{Latitude}
\nomenclature{$Lon$}{Longitude}
\nomenclature{$V'$}{Scaled voltage}
\nomenclature{$i_{ka}$}{Maximum current from and to a bus}
\nomenclature{$i_{rms}$}{RMS current}
\nomenclature{$vm\_pu$}{Voltage Magnitude Per Unit}
\printnomenclature

\subsection{AEMO NEM network:}
AEMO is the Australian Energy Market Operator. Their role is to manage the day-to-day operations of several electricity and gas markets and information services, as well as provide strategic forecasting and planning advice. Specific to this project, AEMO collects data pertinent to the National Electricity Market. The National Electricity Market includes the major transmission infrastructure for the Eastern Seaboard of Australia. Figure 1 shows a graphical representation of this transmission network.

\begin{figure}[h]
\centering
\includegraphics[scale=0.2]{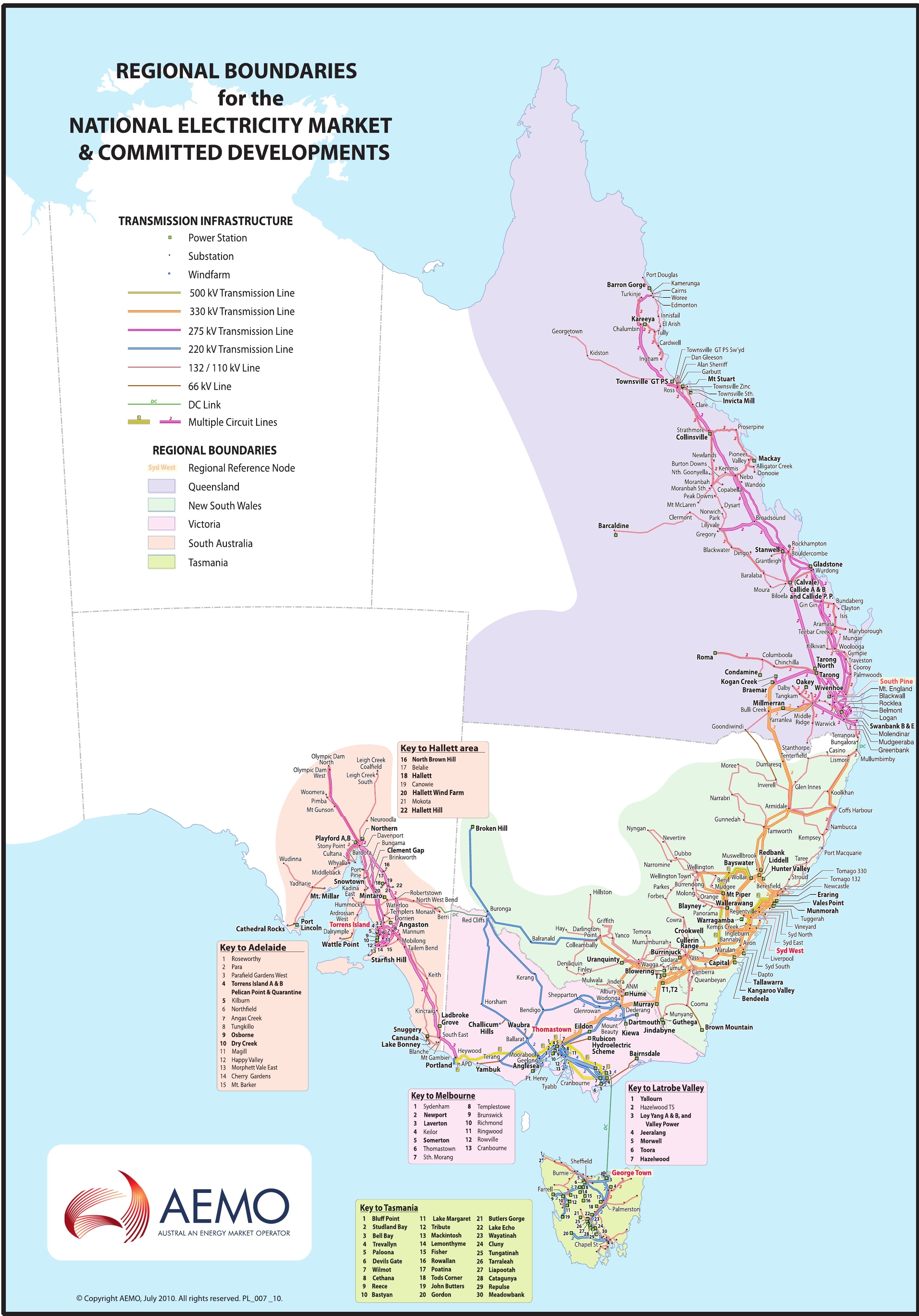}
\caption{AEMO NEM Network (2016). Taken From \cite{NEMNET}. see appendix for larger version}
\end{figure}

\subsection{Transmission and Distribution networks}
The purpose of the electric transmission network is to connect the electric-energy-producing power plants or generating stations with the loads. Most power systems consist of generation facilities that will feed bulk power into a high-voltage bulk transmission network that in turn serves any number of distribution substations \cite{TxtBk2}. The distribution networks in the NEM are not handled by AEMO and are instead contracted out to smaller local companies. These distribution networks are modelled in the NEM network as loads attached at their connection points. More information on this can be found here \cite{TranDist}. A highly simplified network diagram showing a singular connection from the power generator to the substation to the connection point is shown in Figure 2. 

\begin{figure}[h]
\centering
\includegraphics[scale=0.38]{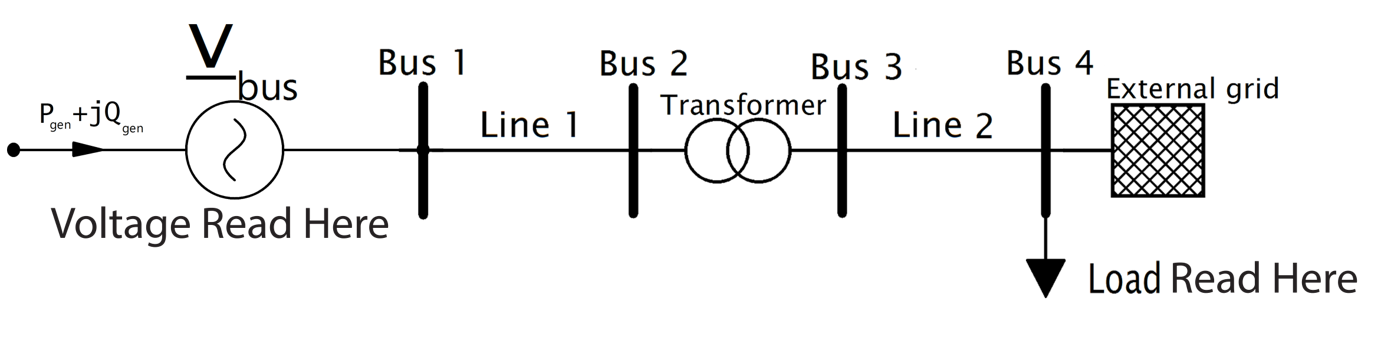}
\caption{Transmission to Distribution diagram.}
\end{figure}

\subsection{Power infrastructure assets}
Four major assets are managed by AEMO. These assets are generators, substations, transmission lines and connection points. Data was acquired from AEMO directly regarding the location and asset information for all of these assets. A graphical representation of these assets can be found here \cite{map}. The general breakdown of the information provided for each asset class is as follows:
\begin{itemize}
  \item Generators:
\end{itemize}
The AEMO NEM network consists of roughly 550 generators. The physical location of each generator was given with geo-data (point coordinates). The maximum and regular generation capacity was provided in MW as well as the maximum and minimum ramp up and ramp down rates. The fuel source was also provided for each generator. These fuel source categorisations include fossil, wind, solar and battery, hydro and Biomass / Waste. A general breakdown of the power demands over 12 months aggregated by fuel mix is shown in Figure 3. Lastly, 
each generator is assigned an asset ID (DUID for generators) which is used as a unique identifier for the generator in the NEM network.

\begin{figure}[h]
\centering
\includegraphics[scale=0.2]{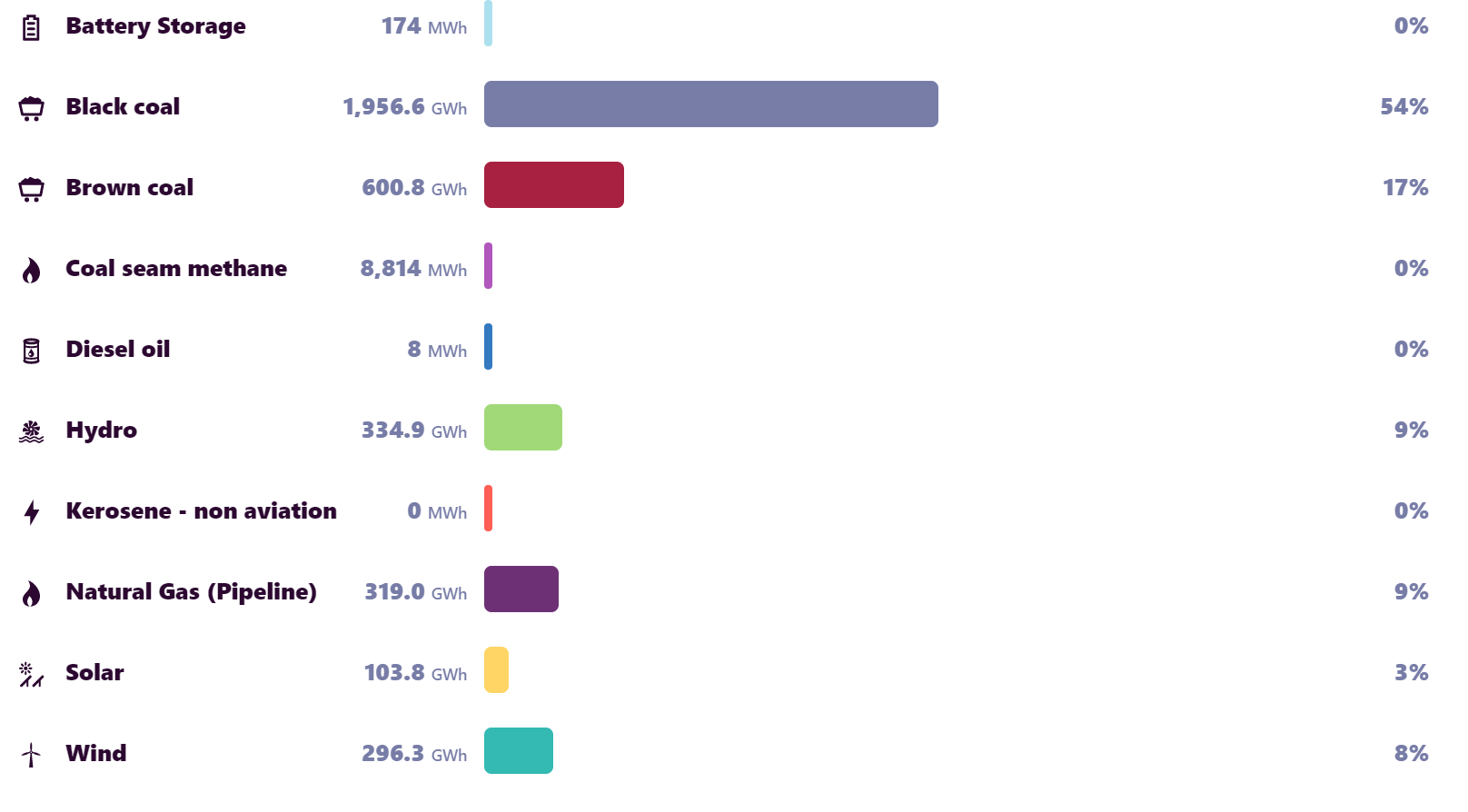}
\caption{NEM power generation broken down by fuel source acquired from  \cite{fuel_mix}}
\end{figure}

\begin{itemize}
  \item Substations:
\end{itemize}

The AEMO NEM network consists of roughly 700 sub-stations. The physical location of each substation was given with geo-data (point coordinates). The high and low voltage power rating for each substation was provided as well as each assets name. 

\begin{itemize}
  \item Connection point:
\end{itemize}

The AEMO NEM network consists of roughly 350 connection points which model higher-level distribution networks. The physical location of each connection point was given with geo-data (point coordinates) An asset ID was also assigned to each connection point (CPID for connection points) which is used as a unique identifier for the generator in the NEM network. 

\begin{itemize}
  \item Transmission lines:
\end{itemize}

The AEMO NEM network consists of roughly 1700 transmission lines. The physical location of each transmission line was given as a geometry collection which consisted of a set of co-ordinate points spanning across the length of each transmission line. The capacity of these transmission lines was given in kilo-volts. 6 sizes of transmission lines are present in the NEM network which are:

\begin{itemize}
  \item 500kV
  \item 330kV
  \item 275kV
  \item 220kV
  \item 132/110kV
  \item 66kV
\end{itemize}

All of the assets in the NEM are of Australian quality standard and as such allow for use in a three-phase AC system operating at 50Hz \cite{TxtBk2}. A printout of the assets found within the NEM network is shown in Figure 4. This printout uses the geo-data provided by AEMO to place the assets. It can be observed from Figure 4 that the data provided by AEMO is not complete as the transmission line infrastructure is missing from numerous generators and substations. It should also be noted that the geo-data provided by AEMO does not match exactly to connect the transmission line infrastructure to the remainder of the NEM assets.   

\begin{figure}[h]
\centering
\includegraphics[scale=1]{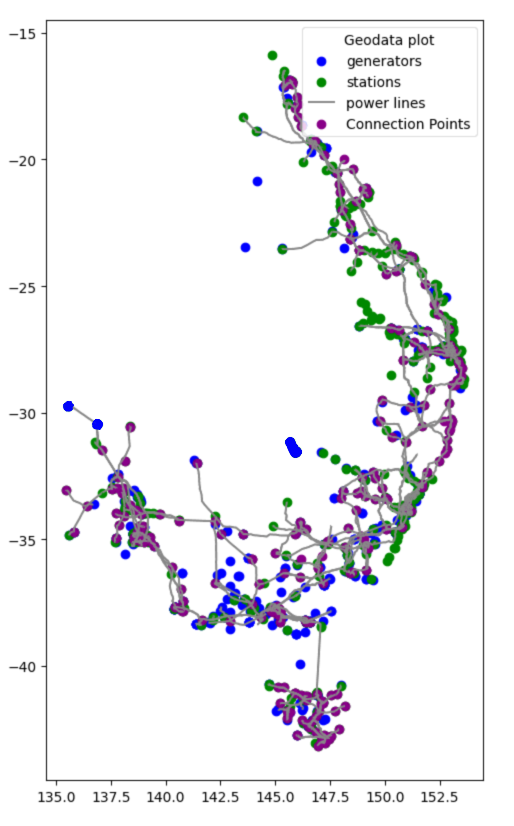}
\caption{AEMO NEM assets. See appendix for larger version}
\end{figure}

\subsection{Australian Power infrastructure plans}
Presently,  ~80\% of the power provided by the NEM comes from fossil fuels. As laid out in \cite{ISP} the power generation infrastructure is set to experience a phasing out of coal fire power stations as 15GW (instantaneous power penetration)  or 63\% of Australia’s coal-fired generation that will reach the end of its technical life and so likely retire by 2040. These major power generation assets are set to be replaced by renewable energy power generation assets due to the falling costs of renewable electricity resulting in renewable electricity being competitive with fuels per unit energy in the replacement process \cite{renewables}, \cite{AEMO_renewables}. Since renewable power operates intermittently, over 30GW of new grid-scale renewables is needed to replace the 15GW of coal-fired power stations being retired. To achieve this AEMO has proposed 5 scenarios for the phase-in of renewable power generation assets with allocated renewable energy zones for these assets. More information on these plans can be found here \cite{AEMO_renewables}. As a \textbf{very} broad oversimplification of these plans, AEMO is moving towards a high instantaneous penetration of wind and commercial solar (as distributed commercial networks) which will alleviate the need for larger generation assets. Some issues may become more prevalent as the need for base-load power deteriorates forcing larger stations to turn off their generators. Issues may also arise from a lack of inertia in the overall system due to the replacement of large generators with a sparsely connected distributed network. Nevertheless, the current plan is to move towards wind and solar as AEMO plans to have 27GW of wind and solar (or 75\% instantaneous penetration) in the NEM by 2020 through their central plan. A heavy reliance on the majority of solar and wind power generation in the NEM is rather risky due to the storage requirements of both (presently, all the battery storage facilities in the NEM account for roughly 0\% of instantaneous power penetration). The AEMO network is relying too heavily on the majority of Australia's power demands are commercial and will peak during the day when renewables will be sufficient to power the NEM. A more robust design/plan would be to incorporate hydro and hydro storage (as is being done on a relatively small scale with the Battery Of The Nation (BOTN) project). Massive success is observed around the Canberra region (due to Snowy Hydro) and In Tasmania wherein a majority of energy is produced through Hydro and instantaneous penetration is relatively stable throughout the day. (this can be observed in real-time through  \cite{fuel_mix}, fuel mix tab, current. The same data for this is also found here \cite{data_model})

\subsection{Power Flow calculations}
A load flow analysis allows the identification of real and reactive power flows, voltage profiles, power factor and any overloads over a network and its assets \cite{TxtBk1}. These functions were developed to automate power system controls to optimise power generation about real-time demand with the end goal being a system that allows a security-constrained scheduling calculation to be initiated, completed and dispatched to the power system entirely automatically without human intervention \cite{TxtBk3}. When an electrical system (such as the NEM network ) operates within the bounds of its steady-state mode, the basic calculation required to determine the characteristics of this state is termed load or power flow. This load flow analysis is generally conducted through optimal power flow calculations wherein there is a need to iterate with contingency analysis. The objective of this iterative analysis is to determine the steady-state operating network characteristics such as the magnitudes and angles of the real and reactive power flows in the individual transmission assets, losses and the reactive power generated or absorbed at voltage-controlled buses for a given set of bus-bar loads which will generally consist of variables such as the load magnitude described bu its real and reactive power \cite{TxtBk3}. When developing a load-flow solution method there are generally 5 main properties that must be considered the computational speed and storage required to find a solution, the solution reliability, versatility and simplicity. The type of solution required will also affect the algorithm used to solve a set of power flow variables i.e. whether the solution is required to be single case or multiple case, accurate or approximate etc. \cite{TxtBk2}

\subsection{Network Modelling}
The assets in a network are modelled by their equivalent circuit. That is, the characteristics describing the assets (e.g. inductance, capacitance resistance etc) are used. Once the assets of a network have been constrained to nodes within some computational software, they can be described such as to comply with Kirchhoff's Laws and can then be solved with methods such as mesh and nodal analysis. For this project, the 4 assets that have been modelled are generators, substations (or transformers), loads and transmission lines. As an example, the transmission lines of the network when modelled are generally done so with regards to the total resistance and this inductive reluctance being modelled in the series arms and the total capacitance being represented in the network shunt arms (although, in PandaPower the conductance of the line per KM is used to describe the line)\cite{TxtBk3}\cite{PandaPowerLine}. 

\subsection{Power Flow Algorithms}
As alluded to earlier simple methods such as mesh and nodal analysis can be used in an iterative form to find solutions to power flow problems through a set of conditioned matrices. Generally, the real and reactive power of the mesh ($P_k, Q_k$) will be used to solve the voltage magnitude and phase angle ($V_k, \theta_k$). For a voltage-controlled bus, the total real power ($P_k$) is specified, and the voltage magnitude ($V_k$) is maintained at a specified value by reactive power. For a non-voltage controlled bus, the total power ($P_k+ jQ_k$) is specified and for a slack or swing bus, the system losses are not known and as such the total power can't be specified. This bus is usually assigned as the systems phase reference and as such the complex voltage of this bus ($E_s = V_S \angle \theta_s$) must be specified. \cite{TxtBk1}\cite{TxtBk2}\cite{TxtBk3}. 
The power at a particular bus will then be given by:
\begin{equation}
S_k = P_k+ jQ_k = E_kI^*_k = E_k\sum_{m\epsilon k} y^*_{km}E^*_m
\end{equation}

This can also be described in terms of the current balance at each bus:
\begin{equation}
    \sum_{m\epsilon k} y_{km} V_m + Y_{kV_k} = \frac{S^*_k}{V^*_k}
\end{equation}

where $y_{km}$ are the elements of the transmission admittance matrix, $Y_k$ are shunt admittances and the index m runs over all buses within a network including the swing buses. Focus is placed on the voltage-controlled (PV) buses in the network. The magnitude of a bus ($|V_K|$) is kept constant using a variable injection of reactive power ($Q_K$) by the generators in the network This turns the RHS of the above equation non-linear and multi-valued. The load flow algorithm will then be used to solve a set of simultaneous nonlinear algebraic power equations specified by either nodal or mesh analysis (solving the conditioned matrices determined through system parameters such as Transmission line and generator parameters etc.) and will solve for any unknown variable. This is generally done through iterations so that a solution can be converged upon. A Flow diagram for a basic load-flow algorithm depicting this is shown in Figure 5. 

\begin{figure}[h]
\centering
\includegraphics[scale=1]{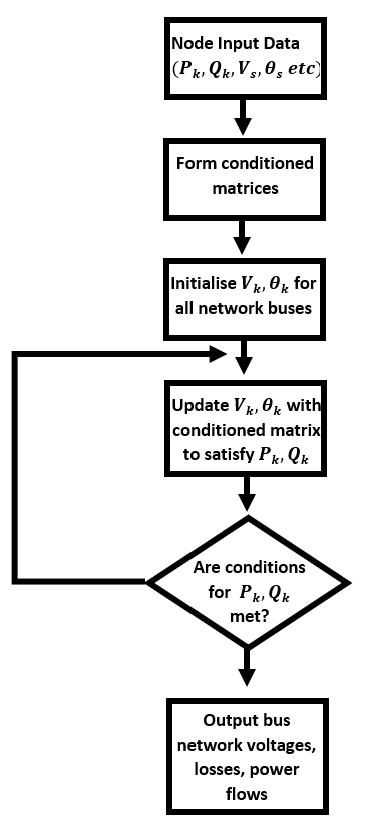}
\caption{Flow diagram for a  basic load-flow algorithm}
\end{figure}

\subsection{PandaPower}
Panda Power is a Python-based BSD-licensed power system analysis tool aimed at the automation of static and quasi-static analysis and optimisation of balanced power systems\cite{PandaPower}. There are many iterative methods used to solve the set of non-linear equations found in power flow algorithms. The 5 that are specified in PandaPower are the Newton-Raphson (nr) Method, Newton-Raphson with Iwamoto multiplier, backward/forward sweep, Gauss-Seidel and fast-decoupled. The most popular method used (and as such the default method used in PandaPower) is the Newton-Raphson method. For a set of simultaneous nonlinear equations ($f_m$) with an equal number of variables ($x_n$) the nr method is described by:

\begin{equation}
    f_m(x_n)=0 \mbox{ for } m=1\xrightarrow[]{}N \mbox{ and } m=1\xrightarrow[]{}N    
\end{equation}

These sets of simultaneous equations will be solved over many iterations. At each iteration of the nr method, the set of simultaneous nonlinear equations is approximated by a linear matrix equation. For each iteration (i), this linearising approximation adds to the above equation an approximation ($x_n^i$) and an error ($\Delta x_n^i$). Therefore,

\begin{equation}
f_m(x_n)=0\xrightarrow[]{}f_m(x_n^i + \Delta x_n^i)=0
\end{equation}
\begin{equation}
\mbox{ for } m=1\xrightarrow[]{}N \mbox{ and } m=1\xrightarrow[]{}N
\end{equation}

Taylor's theorem can then be used to expand the above equation to:

\[f_m(x_n^i + \Delta x_n^i) = f_m(x_n^i) + \Delta x_n^i\cdot f_m'(x_n^i)  + \frac{(\Delta x_n^i)^2}{2!}\cdot f_m''(x_n^i) \]\

 \begin{equation}
 + ... =0
\end{equation}

Assuming that the initial estimate of the set of variables $x_n^i$ is close to the solution for the set of variables, the error $\Delta x_n^i$ will be relatively small. This allows us to neglect the higher powers and gives:
 \begin{equation}
f_m(x_n^i) + \Delta x_n^i\cdot f_m'(x_n^i)=0
\end{equation}

\begin{equation}
\Delta x_n^i = \frac{-f_m(x_n^i)}{f_m'(x_n^i)}
\end{equation}
A Jacobian matrix (J) of first-order partial differential functions ($f_m(x_n^i)$) can then be used to write the equation as:

\begin{equation}
f_m(x_{n^i}) = -J\Delta x_{n^i} \xrightarrow[]{}J_{mn} = \frac{\delta f_m}{\delta x_n}
\end{equation}

This Jacobin matrix represents the slopes of the tangent hyperplanes which approximates our set of simultaneous nonlinear equations formed from the parameters of the network ($f_m(x_n^i)$) at each iteration point.  All elements within the network can be defined with nameplate parameters such as length and resistance for transmission lines, real and nominal power for generators etc. and are internally processed with equivalent circuit models \cite{PandaPower}.

\section{Methodology}\label{Procedure}

\subsection{Creating a network}
The network data obtained from AEMO was all given with coordinate geo-data to place the network assets as seen in figure 4. This cannot be directly ported to a PandaPower network as a PandaPower network requires assets to be linked together through buses. Thus, to use the geo-data provided by AEMO in a PandaPower network a module was written in Python that would link assets together. To achieve a fully connected network, the module connected generators, transformers and loads (connection points) to the transmission line network. This module worked by finding the closest line to an asset (be that generator, transformer or load) and then linking the transmission line to the asset through an interconnecting transmission line. It should be noted here that not a single asset had intersecting geo-coordinates (except for transmission lines that crossed over one another). A simple algorithm was used to find the closest transmission line to an asset. This algorithm was then altered so that transmission lines would intelligently climb up nodes in the network. The new algorithm kept a record of each link that was made between an asset and a transmission line. If the transmission line to asset interconnect connection came in general from the same direction then the first interconnect would be used as a stepping stone to get to the next one. If a cluster of assets were detected in one location then a central transmission line interconnect was created to the centre of the cluster and secondary interconnects were then used to link the network assets. A visual representation of this can be seen in Figure 6.

\begin{figure}[h]
\centering
\includegraphics[scale=0.45]{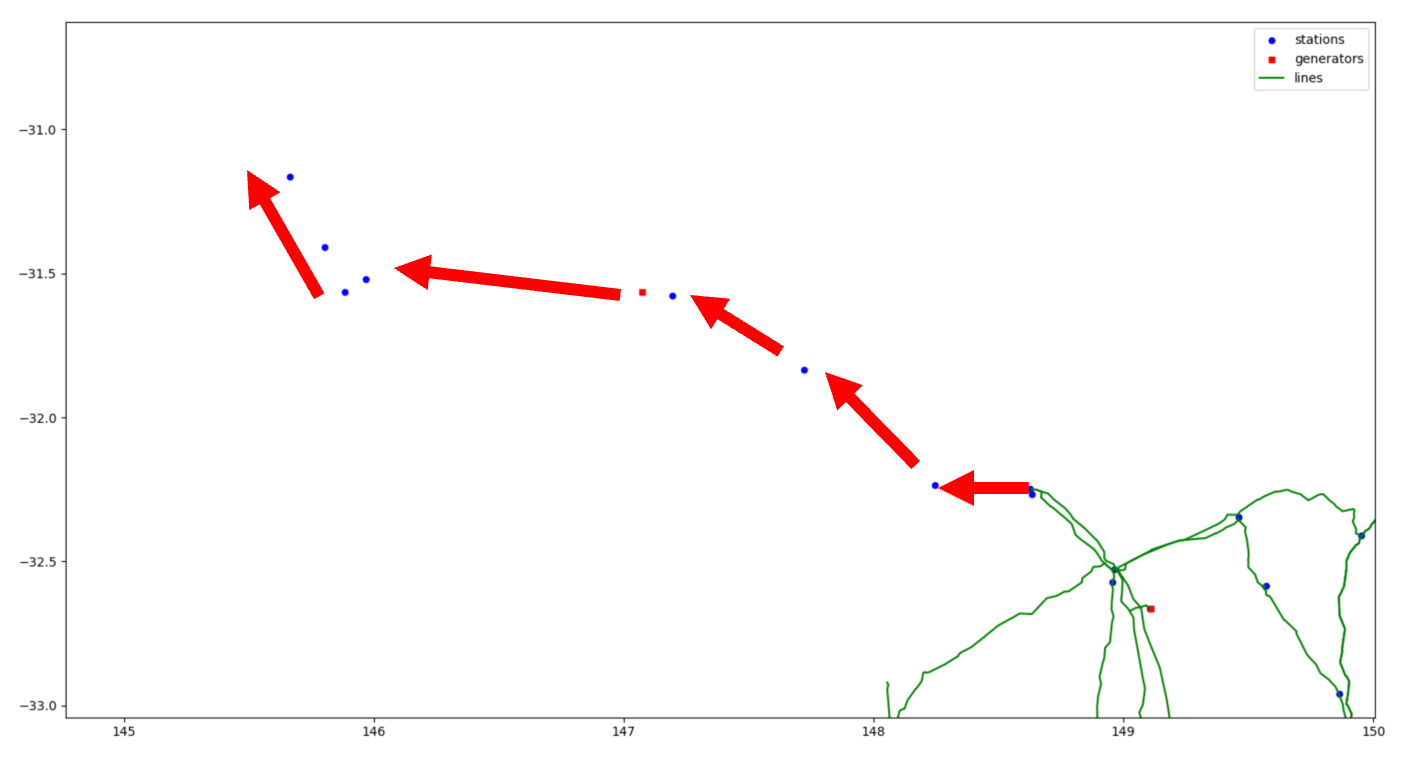}
\caption{Diagram of transmission lines being intelligently added in by climbing nodes}
\end{figure}

With reference to Figure 2, it is not allowed to add two voltage-controlled elements (i.e. Generator and external grid) to buses that are connected through a closed bus-bus switch (or directly connected with only a line), since those buses are fused internally and therefore the same bus leading to convergence issues. This means that a transformer station is required between any generator and the external grid (or load). Theoretically, since the data given from AEMO is a representation of their real network then as long as the above algorithm is implemented as intended this should not be a problem. This was not the case. It was found that clustering in built-up areas (e.g. Syd. West region) coupled with a lack of resolution and clarity from the AEMO data caused linkage errors. As a workaround to this issue, a module was written to detect areas load (connection point to distribution network ) and generator assets were directly connected (similar to that which can be found in the PandaPower diagnostic tool). At points where this was detected the interconnect between the transmission Line and its connection point was moved to the next closest transmission line. 

\subsection{Implementation in PandaPower}
The singular point assets in the AEMO NEM network the generator, transformers and loads(connection points)  were first generated and added to the PandaPower network using the asset information provided by AEMO (e.g. transformer-rated apparent power and iron losses, the rated voltage value of all assets etc). The equivalent circuit diagrams for these assets can be seen in Figure 7. 

\begin{figure}[h]
\centering
\includegraphics[scale=0.7]{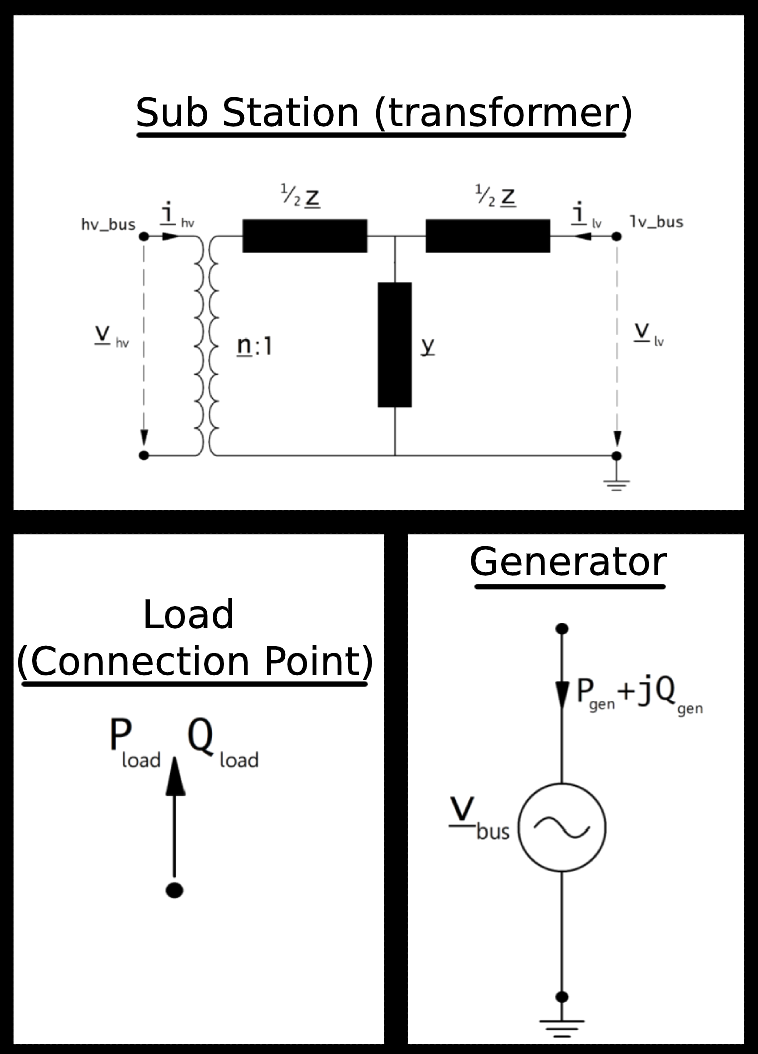}
\caption{Singular point asset equivalent circuit models}
\end{figure}

The values for the real and reactive power ($P_k, Q_k$) for the load and generator assets of the network were initially set to zero and were added in after the entire network was constructed as is discussed further below. As is shown in Figure 7, the transformers were modelled using their t-equivalent circuit (as opposed to their $\pi$ equivalent circuit). Next, all of the transmission lines present in the AEMO asset data set were modelled into the PandaPower network. The distance of each line was required to model a transmission line. The distance between two coordinates was determined using:

\[a = sin^2(\frac{Lat2 - Lat1}{2}) + sin(Lat1)\cdot cos(Lat2) \]

\begin{equation}
\cdot sin^2(\frac{Lon2 - Lon1}{2})
\end{equation}

\begin{equation}
Distance = 2\cdot ArcTan(\sqrt{a}, \sqrt{1-a})\cdot R
\end{equation}

Where a single coordinate point is given by a latitude and longitude pair (lat,lon) and R is the radius of the Earth. This distance equation was used recursively across the entire length of each transmission line coordinate set to determine the full length of each line. Once a line distance was known, other model parameters obtained from AEMO (such as resistance, inductance and capacitance per km etc.) were used to model the transmission line with its $\pi$ equivalent circuit as shown in Figure 8. Also, note that the transmission lines were of course modelled for Australian standards. As such a three-phase, 50Hz power line model was used for each transmission line. Once all of the assets provided by the AEMO data sets had been included into the PandaPower network the node links created from the previous module were used to create the interconnects between all of the assets. The same transmission line model from above was used to model these interconnecting lines in the network. The line parameters were set according to the parameters of the assets for which they were connecting i.e. a 110kV interconnection line would be used to connect a 110kV transmission line to a 110kV generator. If the parameters of the two assets to be linked did not match then the highest required values were used to model the line. The final network (that being, a fully connected PandaPower network) can be seen below in Figure 9.

\begin{figure}[h]
\centering
\includegraphics[scale=0.75]{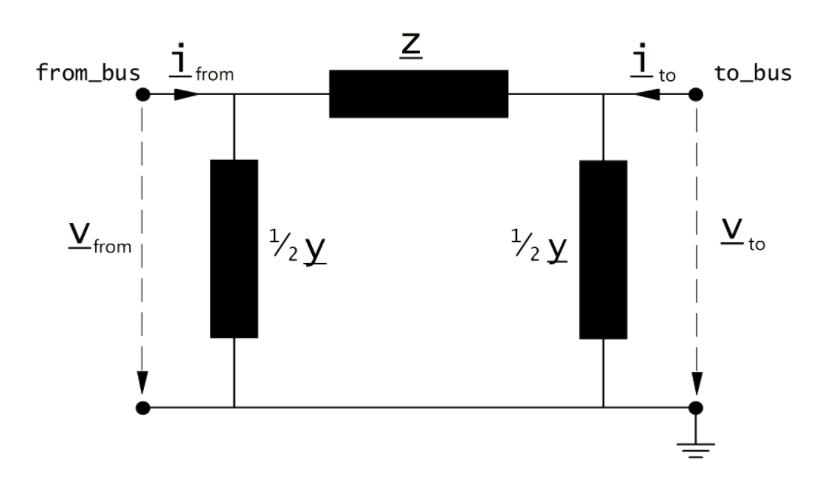}
\caption{Transmission Line equivalent circuit models}
\end{figure}

\begin{figure}[h]
\centering
\includegraphics[scale=0.71]{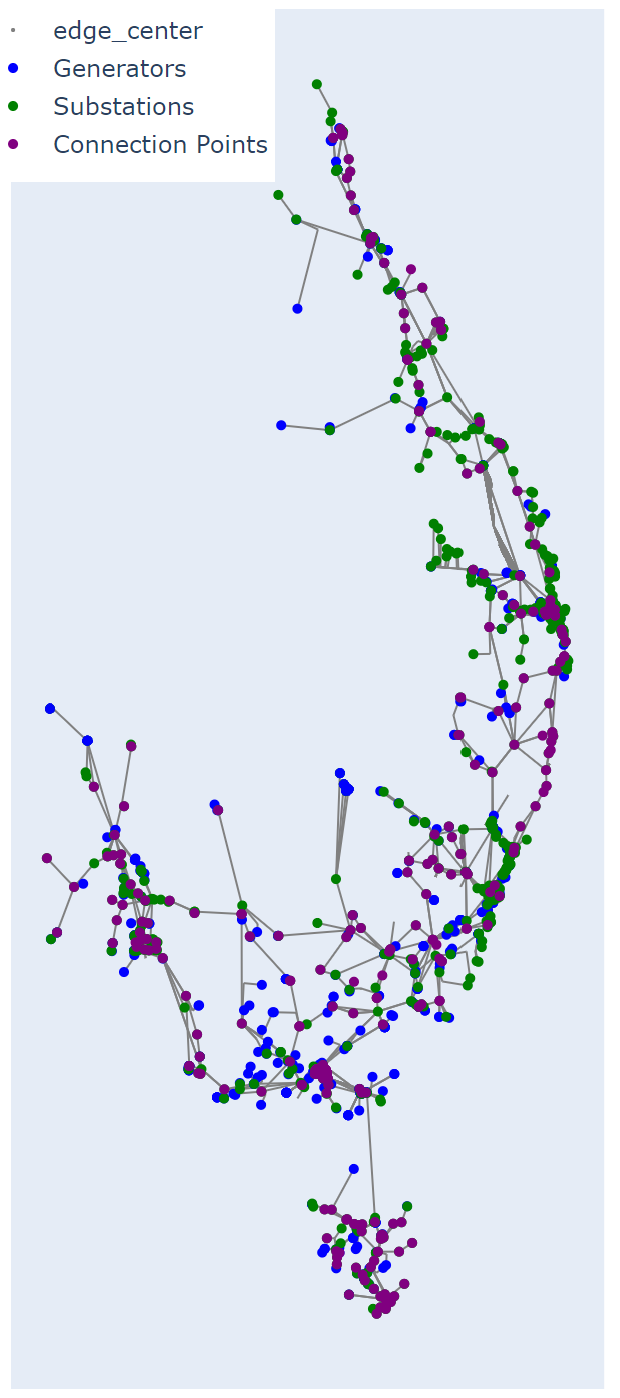}
\caption{Final Printout of Connected PandaPower network. See appendix for larger version}
\end{figure}

\subsection{Integration with generator and load values}
As stated earlier, the values for the real and reactive power ($P_k, Q_k$) for the load and generator assets of the network were initially set to zero. Refer to the introduction section of this report for the importance of these values. This was done so that the network could be set up in such a way that multiple data sets could be run through the network structure. The AEMO data model archive contains actual generation and load data for each generator and connection point in the AEMO NEM network 
which can be found here \cite{data_model} (refer to this website for information on how to use this data \cite{AEMO_GandL}). The data in this archive contains readings across the whole NEM network at 5-minute intervals for operational load demands and generator load dispatch. This data is linked to the assets in the NEM through the assets network unique identifier. Only the data for 2020 was used for this project. A module was thus written to alter the existing PandaPower network generated from the previous module so that the real and reactive power of the generators and loads in the network using their generator ID connection point ID (DUID and CPID)

\subsection{Issues on Convergence}
The PandaPower network developed from the AEMO NEM network was observed to be a primarily radial system in that the path for power to flow from its source (generators) to each customer (connection points) was singular and did not loop back around as it would in a loop or networked system. This makes sense as a radial distribution system is the cheapest to build, and is widely used in sparsely populated areas \cite{TxtBk1}. An example of the radial network infrastructure type that is generally observed in the NEM network is shown in Figure 10 which shows the Cairns region. 

\begin{figure}[h]
\centering
\includegraphics[scale=0.25]{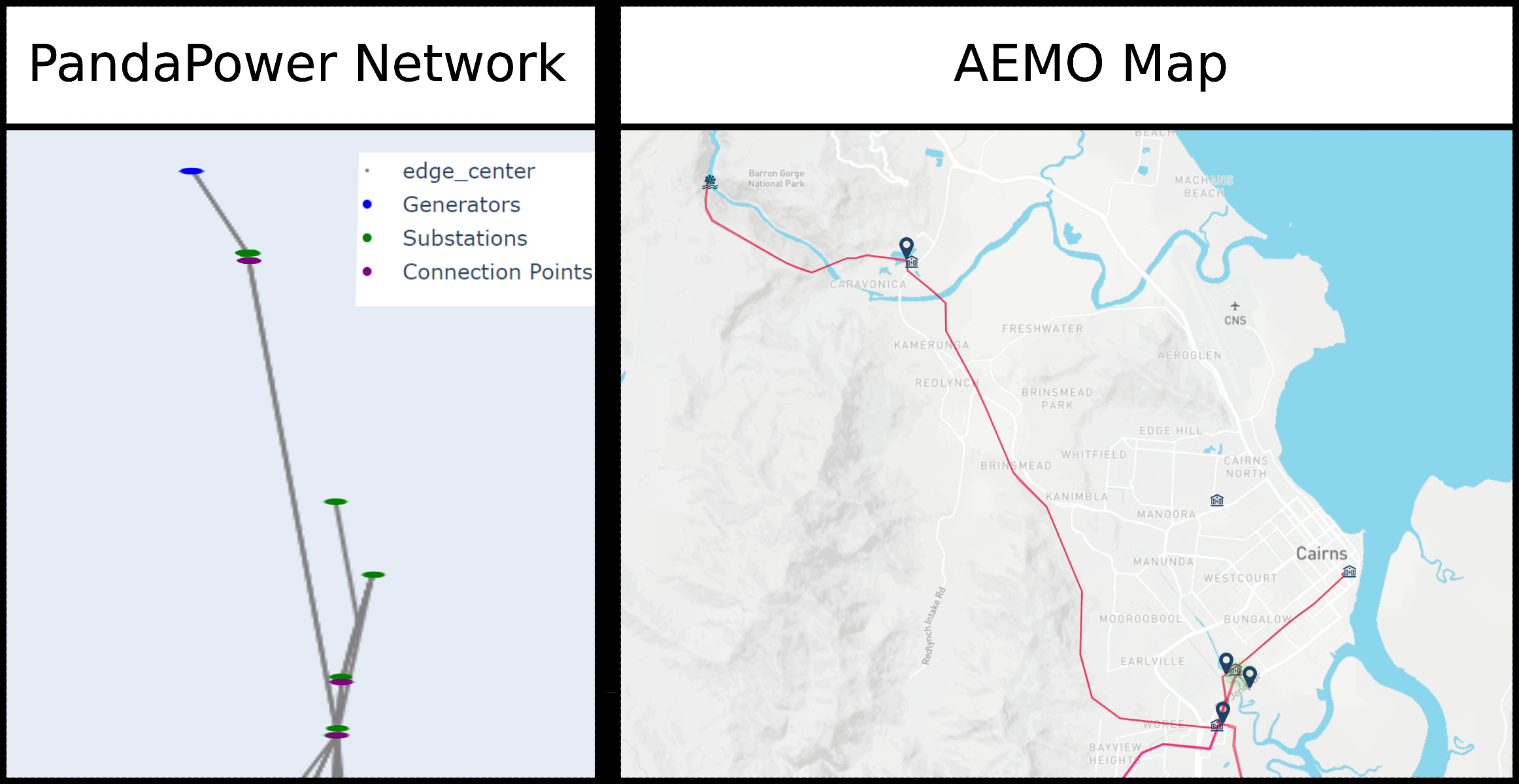}
\caption{Example of the radial network that is generally observed in NEM network. Left is the PandaPower printout. Right in the same area but taken from the AEMO interactive map \cite{map}. See appendix for larger version}
\end{figure}

From Figure 10 we can observe that the Barron hydro 30MW generation station (shown to the north) feeds directly into the Kamerunga power station which then feeds into the Kamerunga connection point ( which predominantly supplies residential, commercial, and industrial purposes). The Barron hydro generation station is then also connected to Cairns City through a 132kV power line. The two power stations to the south (T51 Cairns and H39 Woree) are fed from Coal-fired power stations (not shown as they were too far away). These power stations then feed directly into three connection points the Woree distribution network for the Woree station and the Cairns and Cairns City distribution networks for the Cairns substation, all three of which predominately feed residential and commercial purposes. On a side note, these coal-fired power stations provide roughly 85\% of the power for Cairns (which can be seen through the load demand of the connection points and the generation load of incoming power plants, check \cite{map} for more) and will generally experience a loss of roughly 40\% due to the distances needed to transport that power \cite{cairns_power}. Regardless of where the power comes from though it can be seen that the transmission system observed here (as it is observed in most of the NEM) is a large Generation station feeding directly into a sub-station which will then feed directly into a connection point with no mesh and or loops which is characterised by a radial system \cite{TxtBk1}. This observation has also been confirmed by AEMO here \cite{Radial1} and here \cite{Radial2}. Solving for the optimal power flow in a radial network is often not possible due to the non-convex feasible set of the network which generally leads to a static voltage instability condition that will in turn lead to convergence issues for the solver \cite{convergence_1}, \cite{convergence_2}, \cite{convergence_3}. The key takeaway from this is that the AEMO NEM network could not be solved using the traditional optimal power flow solvers (such as the Newton-Raphson solver described earlier). PandaPower does offer an algorithm that is specially suited for radial and weakly-meshed networks being the backward/forward sweep but this algorithm was also not strong enough to solve for the AEMO NEM network conditions and also ran into convergence issues. A major issue was hence identified in this project - The AEMO NEM network had been simulated in PandaPower but PandaPower did not offer an algorithm that could adequately solve the radial and weakly-meshed network. An investigation into other power flow solvers (e.g. GridCal, MATPOWER (and PYPOWER by extension), PSAT etc) found that no freely available software suites featured algorithms that were adequately suited to solve this network topology. There are many solutions for optimal power flow problems for radial networks (as this is quite a common problem - most power networks are radial). For example, a second-order cone program is suggested here \cite{convergence_2}, Conic programming is suggested here \cite{convergence_3} and an unbalanced three-phase power flow method is suggested here \cite{convergence_1}. These solutions (and many more like them) are not however backed up by a supporting library and would require extensive research and time to implement correctly. It was then suggested to me by the PandaPower development team that I investigate the HELM which did have some Python library support.

\subsection{Use of HELM algorithm}
The Holomorphic Embedding Load-Flow Method (HELM) solver utilises complex analysis to solve optimal power flow problems of networks giving bus parameters as rational approximants. As stated in \cite{HELM2} the HELM solver is guaranteed to find a high-voltage (operable) solution find solution if it exists. Regarding equations 1 and 2 found above, the problem of optimal power flow consists of solving for the node voltages $V_k$ for a given set of power injections $S_K$. The issue with the traditional methods used to solve these simultaneous equation problems is that they rely too heavily on numerical iteration as the root-finding technique and as such the choice of an initial seed will often determine whether the method is successful in finding a solution \cite{HELM1}. The HELM method eliminates the need for an initial seed by computing the formal power series corresponding to the desired solution and then determining the numerical solution through the use of the Pad´e approximants. \cite{HELM1}. The HELM method takes the problem seen in equation 2 and computes its projective invariance through scaling the voltages as:

\begin{equation}
    V_k^{'} = \lambda V_k \xrightarrow[]{} \sum_{m \epsilon k} y_{kmVm^{'}} = |\lambda|^2 \cdot \frac{S^*_k}{V^{'*}_k}
\end{equation}

A complex parameter ($s=|\lambda|^2$) is then embedded into equation 2 such as to allow work to be done in the complex domain, where the fundamental theorem of algebra guarantees that all zeros exist and allows the use of complex techniques predominately derived from holomorphicity \cite{HELM2}. Equation 2 can thus be transformed to:

\begin{equation}
    \sum_{m\epsilon k} y_{km} V_{m(s)} =s \cdot \frac{S^*_k}{V^*_k(s^*)}
\end{equation}

Which is equivalent to equation 12. 
This equation must then be expressed as a set of mirror images due to the nature of complex conjugates with regard to holomorphicity preservation:

\[
    \sum_{m\epsilon k} y_{km} V_{m(s)} =s \cdot \frac{S^*_k}{\hat{V}_k(s)}
\]

\begin{equation}
    \sum_{m\epsilon k} y^*_{km} \hat{V}_{m(s)} =s \cdot \frac{S_k}{V_k(s)}
\end{equation}

Note here that $\hat{V}_k(s)$ and $V_k(s)$ are two sets of independent holomorphic functions. Buchberger’s algorithm on Gr$\ddot{o}$bner basis theory can then be used to describe the system (14) as a polynomial equation:

\begin{equation}
    B(s,V_1) = \sum_{n=0}^{deg B} a_n(s)V_1^n = 0
\end{equation}

A solution to the variable $V_1(s)$ can then be found from equation 15 which in turn can be used to solve for $V_m(s)$ and $\hat{V}_m(s)$ which can then be used to solve for the bus variables $V_k$ and $\theta_k$ all through simple back substitution. The use of the complex domain and the sophisticated analytical tools found within allows for the HELM method to be a much more robust optimal power flow solver which is much more suited to solve radial networks. The implementation of the HELM method was then adapted from the work done by \cite{HELM2} for use in PandaPower and used as the algorithm to solve the power flow problem set by the PandaPower network model of the NEM network developed earlier.

\section{Results}\label{results}
The solution to the unloaded PandaPower network (wherein the bus power values are set to zero) was first evaluated using the HELM solver. The results for this power flow resolution are shown in Figure 11. Two data sets were then run through the PandaPower network and the HELM solver was used to find a solution to the power flow problem. In both sets of data, the generator and load values were averaged over some time for January and August of 2020 (data available in the NemWeb archive). This was done for peak usage time (1700-2100) and the middle of off-peak usage time (2300-0400). The results for peak usage time are shown in figure 12 and the results for off-peak usage time are shown in figure 13.

\begin{figure*}
\centering
\includegraphics[scale=1.1]{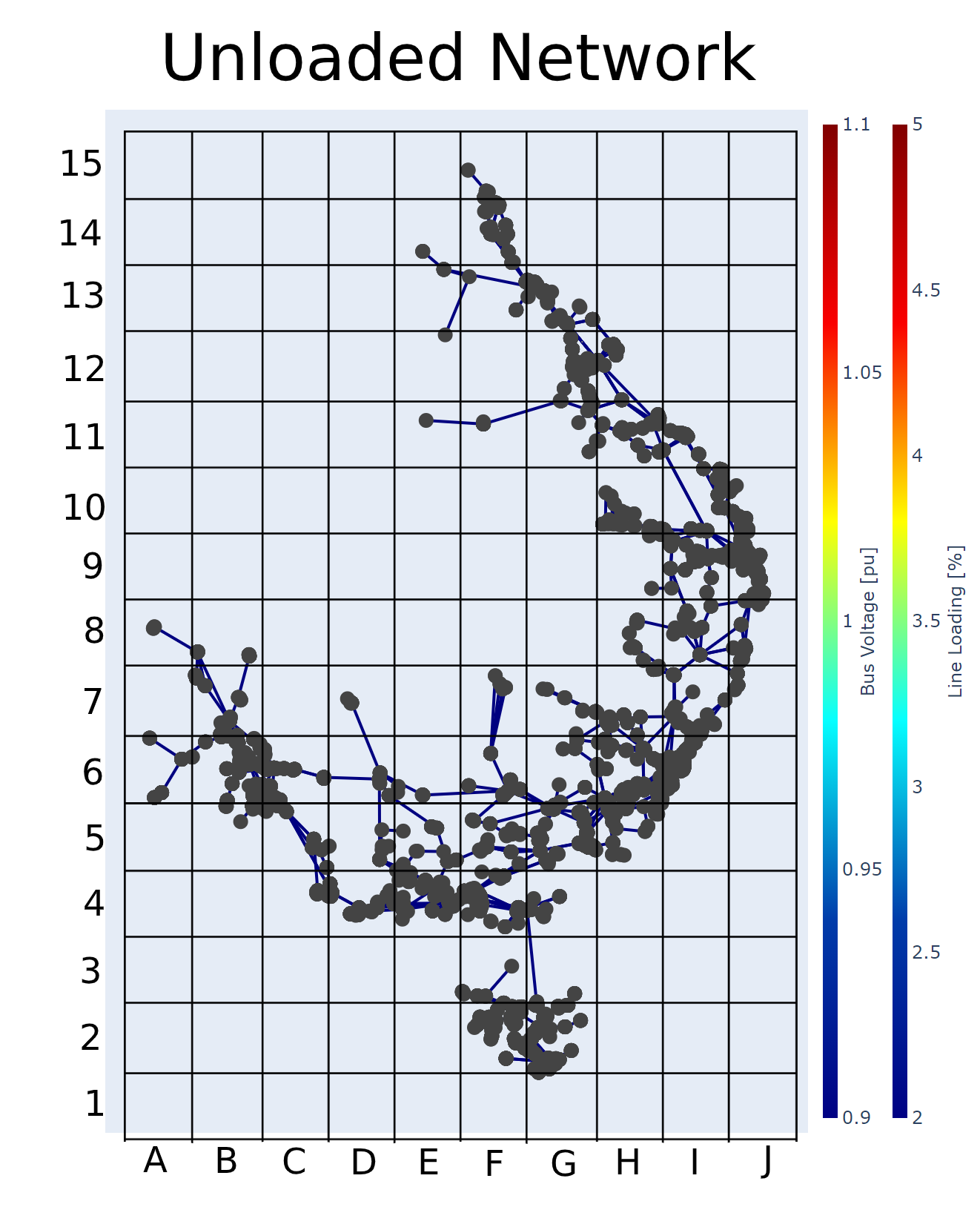}
\caption{HELM power flow solution for unloaded AEMO for NEM network}
\end{figure*}

\begin{figure*}
\centering
\includegraphics[scale=1.1]{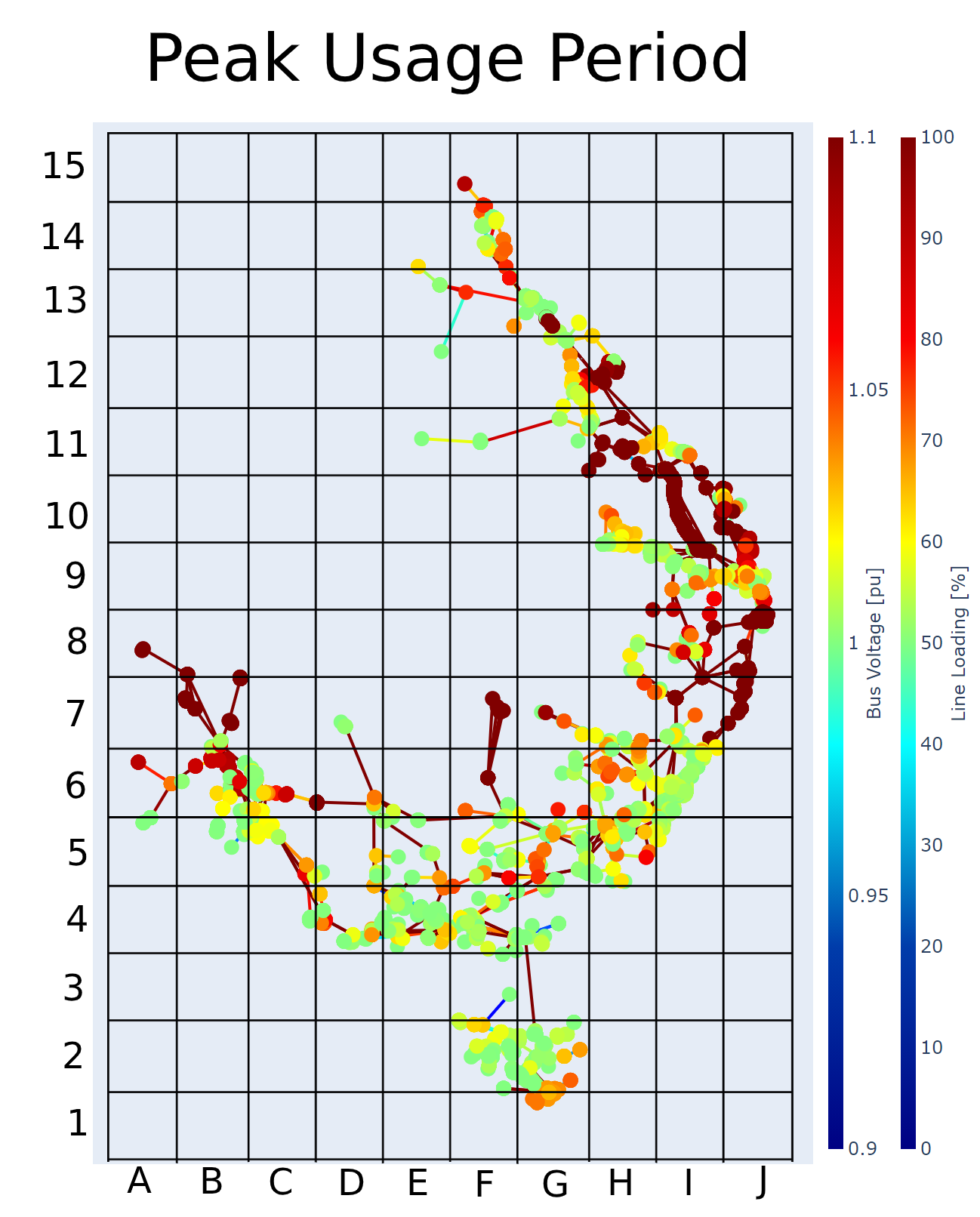}
\caption{HELM power flow solution for peak usage time for AEMO NEM network}
\end{figure*}

\begin{figure*}
\centering
\includegraphics[scale=1.1]{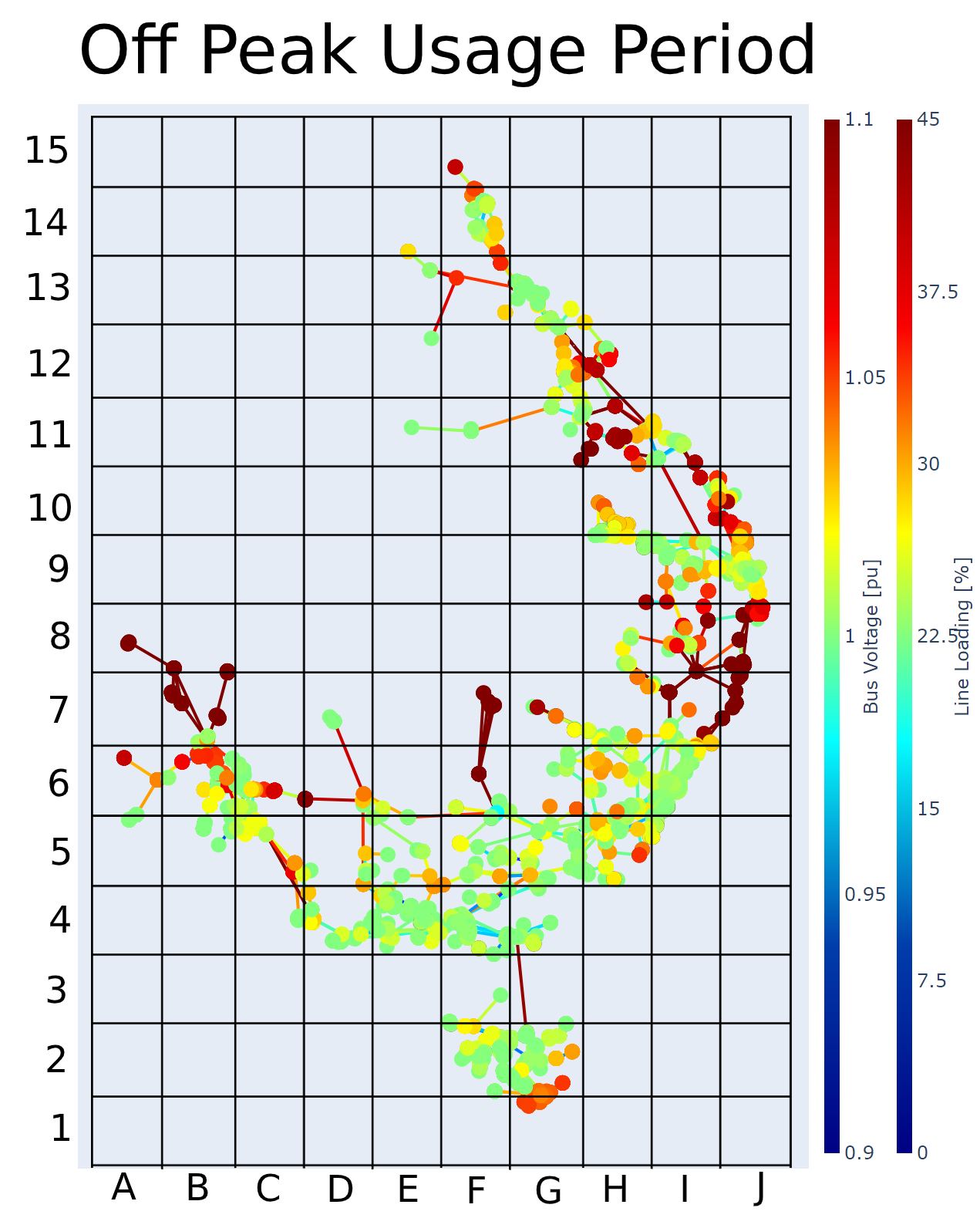}
\caption{HELM power flow solution for off-peak usage time for AEMO NEM network}
\end{figure*}

\section{Discussion}\label{Discussion}

\textit{Line Loading Results}
The line loading result shown in the three results figures describes the current within the wire as a percentage of the maximum current load allowed through the wire (which is specified by the network parameters) That is, 

\begin{equation}
    \mbox{Line Loading = } \frac{i_{ka}}{i_{max k_a} \cdot df \cdot parallel} \cdot 100
\end{equation}

The load flow through these transmission lines will depend upon many factors of the associated system such as the voltage magnitudes, angles and active and reactive power. There are many negligible sources of loss in transmission lines such as dielectric losses, losses due to parallel untransposed lines etc. The primary source of loss in transmission lines however is due to their resistance. The power dissipated in the form of heat loss occurs as the current attempts to overcome the ohmic resistance of the line. This loss is directly proportional to the square of the RMS current travelling through the line \cite{TxtBk1}:

\begin{equation}
    \mbox{Loss} \propto i_{rms}^2 \xrightarrow[]{}P(loss) = i_{rms}^2 \cdot R
\end{equation}

It can therefore be observed that a line with a higher percentage of line loading will have associated losses. From Figure 11 we can confirm that an unloaded system has no losses through its lines as is to be expected with no power and therefore current travelling through the system. High losses are shown during peak usage periods as can be observed in figure 12. There is specifically high line loading found in grid square (GS)H10. The power line that is located in this area is the 275kV transmission line from Calvale to Tarong. An inspection of the area surrounding this power line reveals that it is to its north connected to two large coal-fired power stations (CALLIDE and CALLIDEC) with a regular power output of 760MW and 840MW respectively (Note that these buses cannot be seen in the figures 12 and 13 due to the connection point and sub-station buses that are placed almost directly on top of them in this area). The 275kV, 384km power line is then connected to substations to its south which further connects to distribution networks in the south. Similar situations are shown in GSB7, GSF7 and GSJ7-8 wherein large power generation stations cause a large line loading percentage in their surrounding transmission line infrastructure. The issue of large percentage line loading (causing a higher loss) is however not observed in regions that utilise a higher percentage of renewable energy resources such as Tasmania and South Australia due to the smaller power outputs of the renewable generation stations that are commonplace in these areas. The same general trend is observed during off-peak usage periods as shown in figure 10. Higher percentage line-loading is still observed around power generation stations which predominately use fossil fuels as their primary fuel source due to the larger power generation output of these power stations. It is however less of an issue during off-peak usage periods as the line-loading percentage across the network is generally half as much during this time. 

\subsection{Voltage magnitude results}
 The voltage magnitude result shown in figures 11-13 describes the magnitude of the voltage at each bus in the AEMO NEM network.

\begin{equation}
    \mbox{Bus voltage = } vm\_pu=|V_{bus}| p.u
\end{equation}

This voltage magnitude is however given as a per-unit system. As such, these voltage magnitudes are expressed as system quantities as a fraction of the defined base unit quantity. In PandaPower these per unit values are relative to the phase-to-phase voltages defined for each bus. That is, the per unit system in Panda Power gives the voltage magnitudes as a fraction of the rated voltage of each bus. 

\begin{equation}
    vm\_pu = \frac{|V_{bus}|Volts}{V_{rated} Volts}
\end{equation}

Figure 11 again confirms that the implemented PandaPower network with HELM solver is functioning as intended as a voltage of magnitude of zero is observed for all buses across the network when it is in its unloaded state i.e. when no power is present in the network, no voltage is present at any bus due to:

\begin{equation}
    P = I\cdot V \xrightarrow{} V = \frac{P}{I}
\end{equation}

Figure 12 shows that during peak time the base-load power stations which can produce power on demand are of course producing a maximal amount of power (see GSA8-B7, F7) which is to be expected for obvious reasons. Figure 13 however shows that this power does not seem to drop during off-peak usage times which is in fact the opposite to what is to be expected from these stations. The area around Griffith (corresponding to GSF7) and Roxby Downs (corresponding to GSB7) was further investigated for an explanation. The power used in these areas was found to be used for predominantly agricultural purposes with some industrial areas present. As stated here \cite{Agr}, the direct use of energy for agricultural needs often includes activities such as Irrigation equipment, Lighting for houses, sheds, and barns, Heating for frost protection in groves and orchards, Heating/cooling of cattle barn, pig or poultry brooder, greenhouse, stock tanks, etc i.e. activities that are relevant at both on and off-peak times which explains the in-variant nature of power usage in these agricultural areas. Examining the difference in areas such as GSG5 which is predominately a solar installation we can observe that power generation drops as would be expected of an installation that is dependent upon sunlight. The in-variant nature of power generation in Tasmania and South Australia should also be noted here and is due to the reliance on wind and Hydro which is time-invariant and can be observed as quite stable when examined over a large enough period (8 months in this case) 

\textit{Power generation VS power loss}
From the above observations, the two major conclusions can be drawn. The first is that hierarchical transmission networks wherein large power generation assets are utilised are inefficient due to the losses observed across the transmission network as a result of line-loading effects. As an example of how large these losses can be the 275kV, 384km power line connecting Calvale and Tarong can be analysed. The line-loading in this transmission line was observed at 100\%. From \cite{loss} the energy lost through transmission lines in a three-phase, 50Hz system is given by:

\begin{equation}
    E = 3R\int_{0}^{T} I^2(t)dt \cdot 10^{-3} = 3I_{rms}^2RT\cdot10^{-3}. 
\end{equation}
The instantaneous power present at the node to the north of this power line is equal to the sum of the buses present here or $P_{bus} = 1600MW$. For a standard Australian, 50Hz 3-phase power line rated at 275kV, the restive property is R = 0.075$\Omega/Km$. The power loss over this transmission line can hence be evaluated to 599.48MW or roughly 37\% of the power output entering the node to the north. A similar result of 40\% power loss is also cited here \cite{cairns_power}. Taking the average spot price for this region of \$85.5 /MWh \cite{lossPrice} and accounting for a 4-hour peak period it can be observed that the monetary loss for this one transmission line due to its high line loading as a result of the large power generation station is:

\begin{equation}
    \frac{\$85.5}{MWh} \cdot  599.48MW \cdot 4h = \$205,022.16
\end{equation}

Or enough money to buy almost 3 Tesla Models S's in just one four-hour period!

\subsection{Missing Data}
From Figure 10 it is clear that there is missing or outdated data in this data set. Two sub-stations are positioned to the west with no clear/direct connection to a connection point or generator. Further evidence of missing data is seen throughout the NEM network printout in Figures 4 and 9. The most visible is the cluster of substations in the middle of Queensland with no connection points and /or generators. In reality, these sub-stations have either been decommissioned or are connected to a distribution network. It should be reiterated here that all data used in this project is real. Real asset location properties have been used to construct the PandaPower network used in this project. Real load and generation data were used to produce figures 11-13. However, whilst all the data used in this project has been real there are clear gaps in the NEM network which has consequences on the solution found by the HELM algorithm. As stated by AEMO, \enquote{AEMO does not guarantee the accuracy of the data or its availability at all}. As such, whilst the simulated network produced is most likely indicative of the real NEM network, it may vary in places from the real network.

\subsection{Future Work and Recommendations}
Further work must be done to fill in the missing data gaps from the data sets provided by AEMO so that a virtual/simulated network can match more closely the real NEM network. Furthermore, the data for rooftop solar PV should be taken into account in further iterations of this project as well as new and upcoming infrastructure projects being undertaken by the power grid companies. In this project, python modules were written to convert the AEMO asset and load and power generation data sets into a PandaPower network. The HELM solver was successfully implemented alongside the PandaPower modules to solve the radial PandaPower network produced. Further modules have been produced so that it is possible to run bulk data through the constructed PandaPower network and HELM solver. The summation of this project is now readily available to be used for further work. Whilst I did not have time to conduct any more work before this project was due (The HELM solver is computationally time-expensive and requires huge iteration amounts to properly converge) the modules are now ready to be taken and used to solve for things such as optimal asset placement. From the results shown above it can be observed that the smaller bus voltage magnitude indicative of renewable generation stations leads to a much smaller line-loading percentage and therefore less losses across the transmission infrastructure in the areas wherein they are observed such as Tasmania and South Australia (which predominately utilise hydro and wind power respectively). It was also shown that the renewable infrastructure used in these areas is generally reliable when their steady-state response is observed. The current renewable integration plan as set by AEMO sees large amounts of solar and wind penetrating the NEM network in small and distributed networks. A more desirable solution to the phase-in of renewable energy generation stations would see the integration of hydroelectric stations into this plan to build upon the success seen in Tasmania. The distributed power generation network is an elegant solution to the line-loading issues seen in areas wherein bulk power is generated from base-load power stations. Through the use of the modules written in this project, a plan can be formulated for the strategic placement of wind, hydro and solar so as to optimise the NEM network and cut down on transmission line losses.

\section{Conclusion}\label{conclusion}
The HELM algorithm was implemented in Python for use as a solver in PandaPower. It was then used in this project to solve the optimal power flow problem introduced by a radial PandaPower. This radial network was generated from the data given by AEMO on the NEM network and included the location and asset information for their generators, substations, power lines and connection points to distribution networks. Large losses were observed in the transmission infrastructure surrounding base-load power plants due to the line-loading present in these areas. These losses were not observed in areas that had a higher percentage of renewables as power generation was generally more distributed in these areas. Furthermore, the voltage levels present at the Hydro and Wind farms across Tasmania and South Australia were found to be stable in their steady state for both peak and off-peak usage times. A distributed network of renewable infrastructure is then proposed as a solution to the issues facing the NEM network. The modules written in this project could be used to determine the optimal placement of the assets for this renewable energy network.

\bibliographystyle{IEEEtran}
\bibliography{references}

\section{Appendix}\label{appendix}

\begin{figure*}
\centering
\includegraphics[scale=0.45]{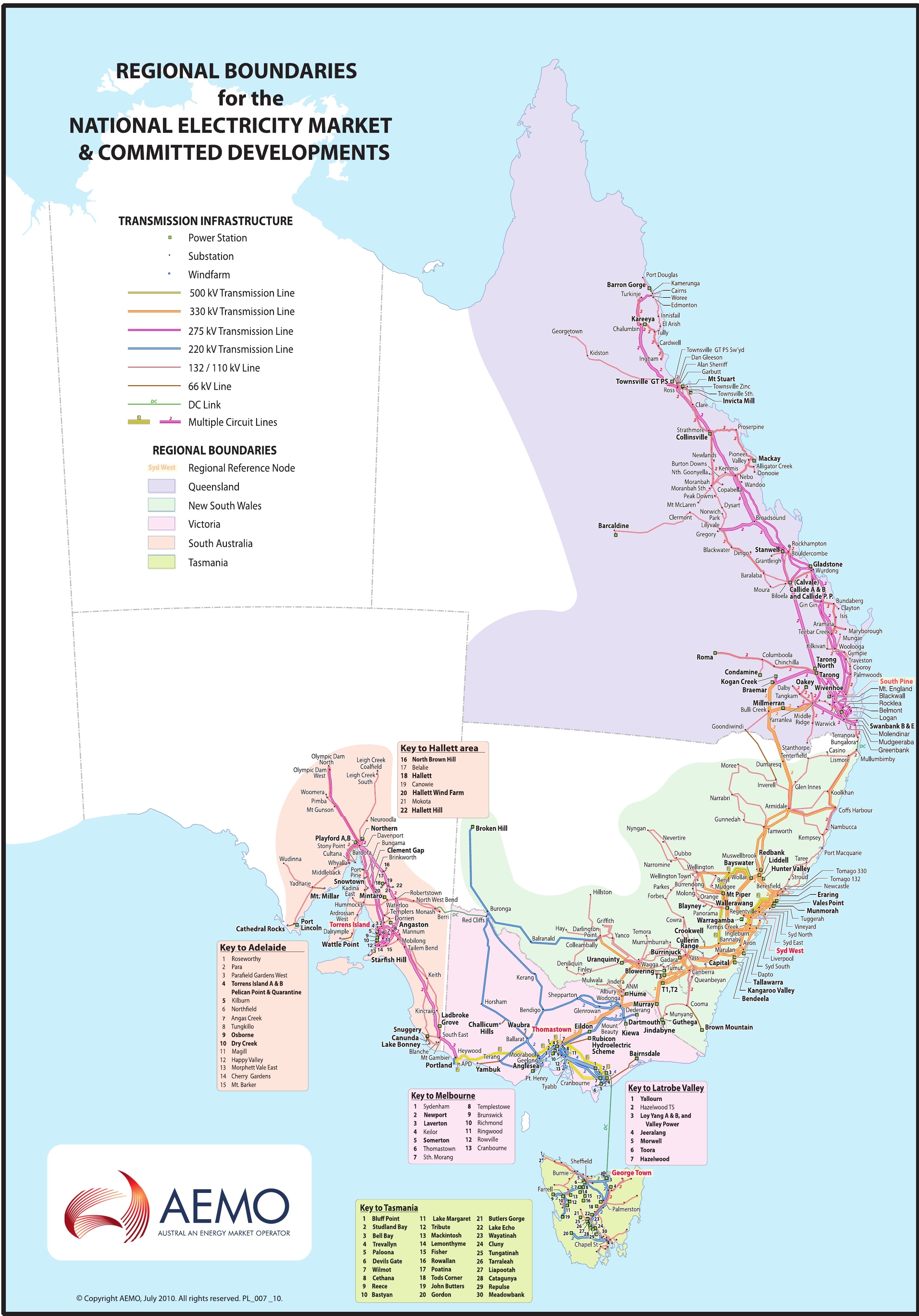}
\caption{AEMO NEM Network (2016). Taken From \cite{NEMNET}}
\end{figure*}

\begin{figure*}
\centering
\includegraphics[scale=2]{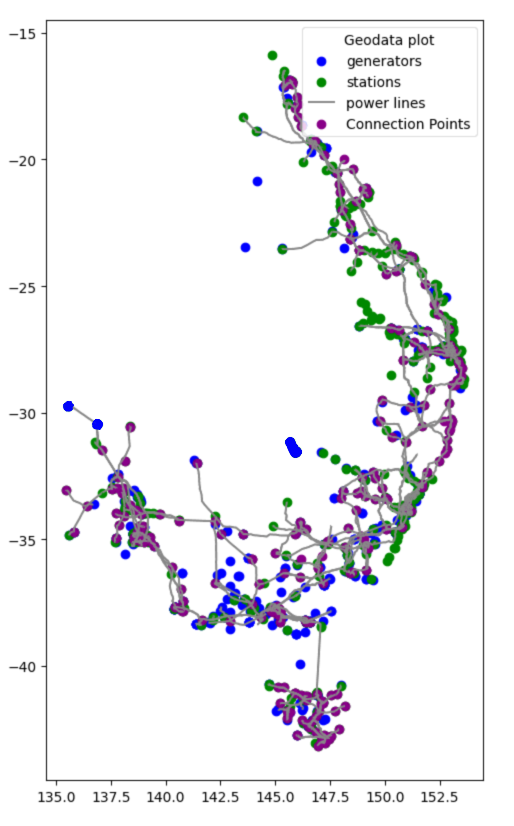}
\caption{AEMO NEM assets.}
\end{figure*}

\begin{figure*}
\centering
\includegraphics[scale=1.25]{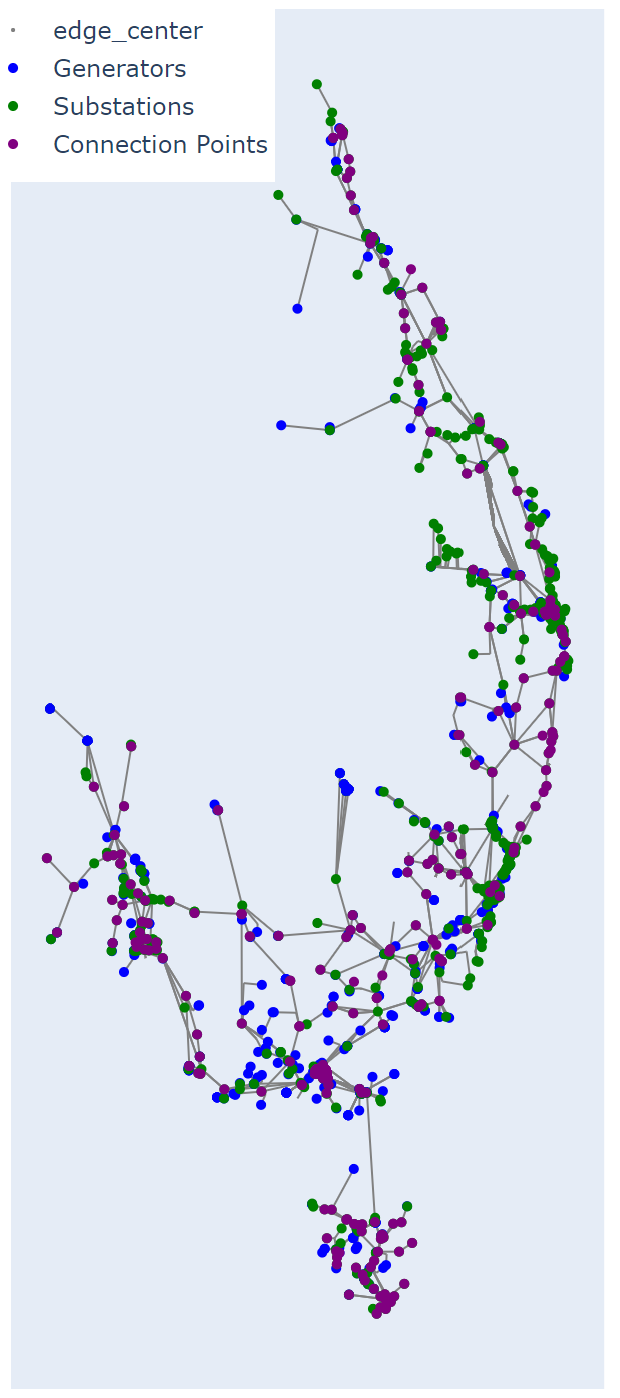}
\caption{Final Printout of Connected PandaPower network. }
\end{figure*}

\begin{figure*}
\includegraphics[scale=0.6]{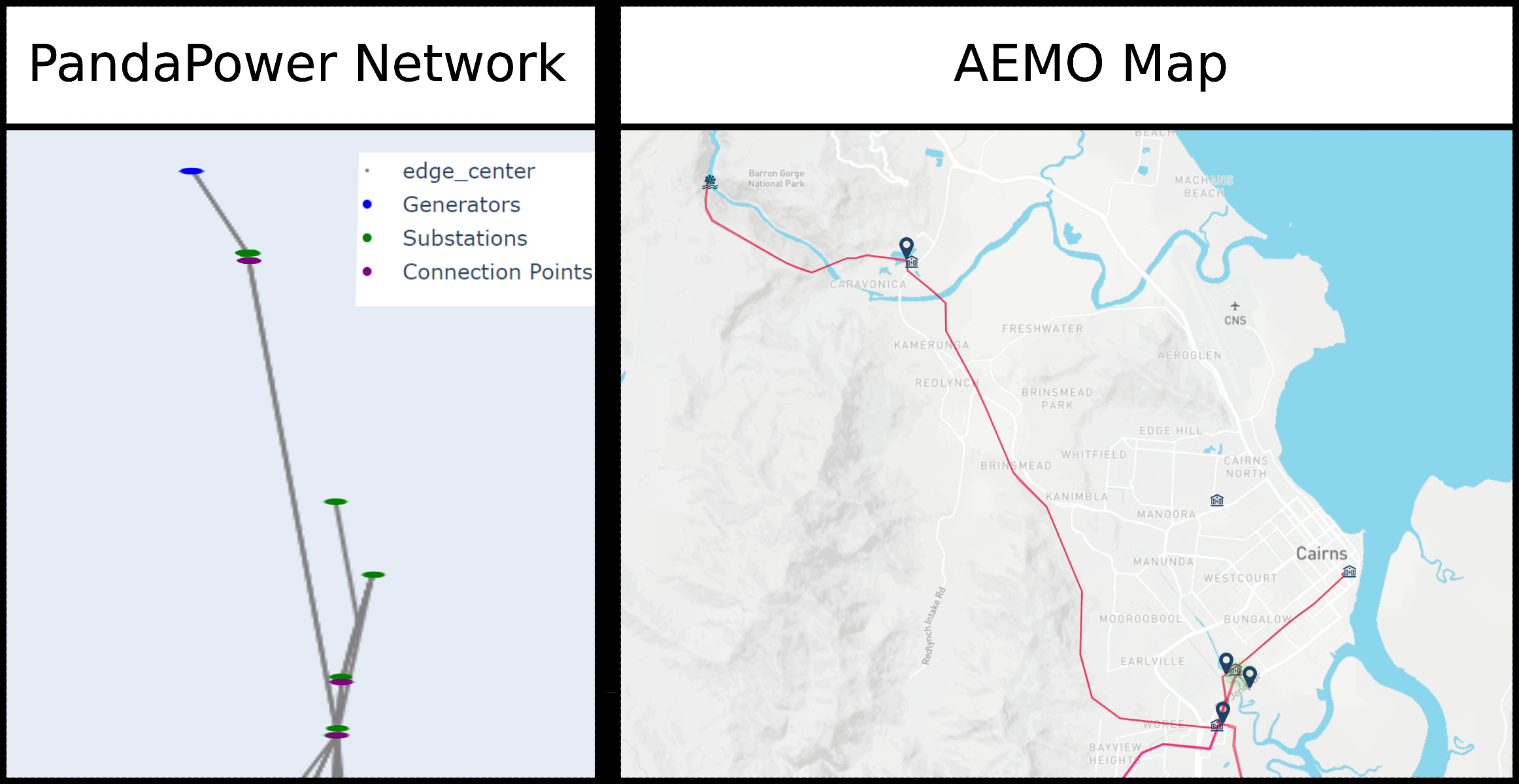}
\caption{Example of radial network that is genreally observed in NEM network. Left is PandaPower printout. Right is the same area but taken from the AEMO interactive map \cite{map}.}
\end{figure*}

\end{document}